\documentclass[reprint, amsmath, amssymb, aps, prl, superscriptaddress, nofootinbib]{revtex4-1}

\usepackage{graphicx}
\usepackage{amsmath}
\usepackage{mathrsfs}
\usepackage{url}
\usepackage{natbib}
\usepackage{color}

\begin{document}

\title{High-Efficiency Cross-Phase Modulation in a Gas-filled Waveguide}
\author{C. Perrella}
 \affiliation{School of Physics, University of Western Australia, Perth, Western Australia 6009, Australia}
 \affiliation{Institute for Photonics and Advanced Sensing (IPAS) and the School of Chemistry and Physics, The University of Adelaide, Adelaide SA 5005, Australia}
\author{P. S. Light}
 \affiliation{School of Physics, University of Western Australia, Perth, Western Australia 6009, Australia}
 \affiliation{Institute for Photonics and Advanced Sensing (IPAS) and the School of Chemistry and Physics, The University of Adelaide, Adelaide SA 5005, Australia}
\author{J. D. Anstie}
  \affiliation{School of Physics, University of Western Australia, Perth, Western Australia 6009, Australia}
  \affiliation{Institute for Photonics and Advanced Sensing (IPAS) and the School of Chemistry and Physics, The University of Adelaide, Adelaide SA 5005, Australia}
 \author{F. Benabid}
 \affiliation{GPPMM group, Xlim Research Institute, CNRS, Universite de Limoges, France}
\author{T. M. Stace}
 \affiliation{Centre for Engineered Quantum Systems, School of Mathematics and Physics, University of Queensland, Brisbane, Queensland 4072, Australia}
\author{A. G. White}
 \affiliation{Centre for Quantum Computing \& Communication Technology,School of Mathematics and Physics, University of Queensland, Brisbane, Australia}
\author{A. N. Luiten}
 \affiliation{School of Physics, University of Western Australia, Perth, Western Australia 6009, Australia}
 \affiliation{Institute for Photonics and Advanced Sensing (IPAS) and the School of Chemistry and Physics, The University of Adelaide, Adelaide SA 5005, Australia}

\begin{abstract}
Strong cross-Kerr non-linearities have been long sought after for quantum information applications.
Recent work has shown that they are intrinsically unreliable in travelling wave configurations: cavity configurations avoid this, but require knowledge of both the non-linearity and the loss.
Here we present  a detailed systematic study of cross-phase modulation, and absorption, in a rubidium vapour confined within a hollow-core photonic crystal fibre.
Using a two-photon transition, we observe phase modulations of up to $\pi$\,rad with a signal power of $25$\,$\mu$W, corresponding to a non-linear Kerr coefficient, $n_{2}$, of $0.8 \times 10^{-6}$\,cm$^{2}$/W, or $1.3 \times 10^{-6}$ rad per photon.
\end{abstract}

\maketitle
%----------------------% Intro %----------------------%
Photons are a promising vehicle for processing \cite{Knill2001, Pryde2003, Chiaverini2004, Browne2005,  Walther2005, Kok2007, Politi2008} and storing \cite{Moiseev2001, Julsgaard2004, Phillips2008, Hosseini2009, Amari2010, Reim2010, Hosseini2011} quantum information.
They are particularly attractive because of their weak interaction with the environment, ensuring long-lived quantum states.
This very feature, however, implies that it is difficult to engineer deterministic interactions between photons, necessitating strong interactions between light and matter \cite{Turchette1995, Milburn1989, Duan2004}.
The best studied light-atom interaction in this regard is the cross-Kerr effect, where an effective interaction between a control and probe field is mediated by a non-linear medium \cite{Milburn1989}.
The interaction is characterised by observing a phase shift on the probe field which varies linearly with the power of the control field.
The largest cross-phase modulation observed to date is $0.2$ rad per photon in microwave waveguides, using a single transmon qubit as the non-linear medium \cite{hoi2012giant}.
At optical frequencies, non-linear optical fibres with cross-Kerr shifts have been directly measured at the level of $10^{-7}$ rad per photon~\cite{Matsuda2009, Matsuda2009APL}.
Recent experiments using vapour-filled hollow-core photonic crystal fibre (HC-PCF) inferred shifts up to $10^{-3}$ rad per photon \cite{Venkataraman2012aa}.
Furthermore, such systems have also been shown to be highly effective all-optical switches \cite{Venkataraman2011, Bajcsy2009}.

Single-pass operation of cross-phase non-linearities is conceptually and technologically alluring.
Recent theoretical \cite{shapiro2006single,fan2012breakdown} and experimental \cite{hoi2012giant} studies have shown that extension to the single-photon regime involves subtleties about the dynamics of the nonlinear medium itself, making extrapolation to the single-photon regime difficult.
Fan et al.\ \cite{fan2012breakdown} showed that for travelling waves, the interplay between quantum noise and the intrinsic saturation of the non-linear medium ensure that single-photon-induced phase-shifts are always too small to be reliably resolved shot-to-shot.
Indeed, data presenting cross-Kerr shifts at optical frequencies have alluded to this being the case \cite{Matsuda2009, Matsuda2009APL}.
This situation can be overcome by embedding the non-linear interaction within a resonant cavity, however, the efficiency of such is dependent on the loss of the non-linear medium.
Hence, the critical physics of this new architecture is captured in the \emph{ratio} of the non-linearity to loss.
Previous work using vapour filled HC-PCF did not address this aspect.
Here we present a systematic study of cross-phase modulation, atomic saturation, and loss for a HC-PCF filled with a rubidium (Rb) vapour.
By demonstrating a large phase-shift with low loss, we open a path to a promising non-cryogenic architecture for scalable quantum information processing. 

%----------------------% Experiment setup %----------------------%
\begin{figure}[b]
	\includegraphics[width=\columnwidth]{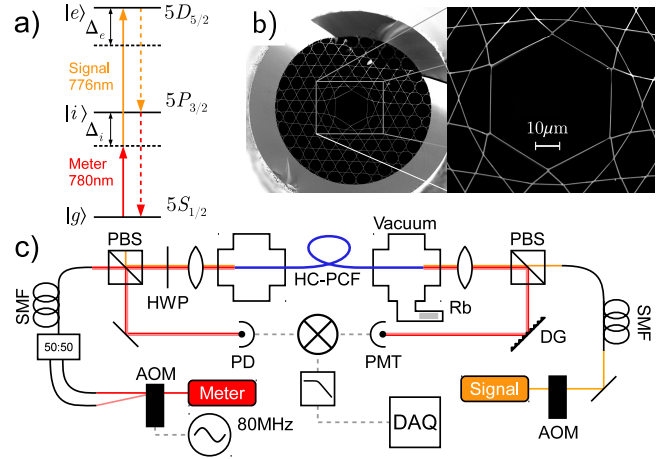}
	\caption{\label{Fig:ExpSetup} a) Energy level diagram of the two-photon transition. Solid arrows are driving lasers, dashed arrows show decay routes. b) Scanning electron microscope image of the kagome HC-PCF being used. c) Schematic of the optical experimental setup. AOM, Acoustic Optic Modulator; SMF, Single mode fibre; PBS, Polarizing beam splitter; HWP, Half wave plate; DG, Diffraction grating; DAQ, Data Acquisition.}
\end{figure}

Coupling between light and a collection of dipoles can be maximised  by matching the transverse dimensions of both optical field and dipoles.
In practice, engineering the atomic dipole moment is difficult, however the advent of HC-PCF enables constriction of the transverse dimensions of the optical field to several microns over arbitrarily long distances \cite{Russell2003, Benabid2009, Benabid2011}.
In our experiment we achieve an extended, and strong, light-atom interaction using a HC-PCF to confine both an optical field and Rb vapour within the fibre's 45\,$\mu$m diameter hollow core~\cite{perrella2013}.
The fibre's kagome lattice cladding---cross-section shown in Fig~\ref{Fig:ExpSetup}b)---provided low-loss guidance from $600$ to $1600$\,nm~\cite{Couny2006}.
The fibre was mounted between two vacuum chambers, one of which contains a dense Rb vapour.
Fluorescence measurements confirmed that over half of the $40$\,cm fibre was filled with Rb.
The Rb density within the fibre was elevated by heating the vacuum chamber and fibre to ${\approx}110$\,$^\circ$C.

The $5S_{1/2} (F\!=\!3) \, {\rightarrow} \,5D_{5/2} (F'\!=\!1\!-\!5)$ two-photon transition of $^{85}$Rb  is used as the basis of the non-linear interaction.
The atomic  energy level scheme, along with decay routes and driving lasers, is depicted in Fig~\ref{Fig:ExpSetup}a).
The two-photon transition strength was resonantly enhanced by the use of a small detuning from the intermediate $5P_{3/2}$ state: this requirement set the wavelengths of the driving lasers at $780$\,nm and $776$\,nm respectively.
For the rest of this paper the ground ($5S_{1/2}$) state will be labeled $|g\rangle$ while the intermediate ($5P_{3/2}$) and excited ($5D_{5/2}$) energy levels are labeled $|i\rangle$ and $|e\rangle$ respectively, with associated rates $\Gamma_{i}$ and $\Gamma_{e}$.
The frequency detuning from the intermediate state is given by $\Delta_i =\omega_{gi} - \omega_{780}$, and the two-photon detuning $\Delta_{e} = \omega_{ge} - (\omega_{780} + \omega_{776})$, where $\omega_{jk}$ denotes the $|j\rangle {\rightarrow} |k\rangle$ transition frequency.

The $|g\rangle {\rightarrow} |e\rangle$ transition was excited using a Doppler-free configuration \cite{Bjorkholm1976}, providing both strong light-atom interaction (absorption ${>}70\%$ for on-resonance pump laser powers ${>}5$\,$\mu$W) and a narrow linewidth ($\Gamma_{e}{\approx}10$\,MHz)~\cite{perrella2013}.
These attributes make this transition ideal for cross-phase modulation experiments, as one can operate at a small detuning which provides simultaneous high interaction but small absorption.
The lineshape of the transition is well described by a Voigt function with a full-width at half-max (FWHM) dominated by transit-time, residual Doppler and magnetic field broadening~\cite{perrella2013}.

Figure~\ref{Fig:ExpSetup}c) shows the optical setup and detection scheme.
The 780\,nm radiation was provided by an extended cavity diode laser (ECDL), while the $776$\,nm radiation came from a Titanium:sapphire laser.
The lasers were coupled into opposite ends of the HC-PCF, enabling Doppler-free spectroscopy of the two-photon transition within the trapped vapour.
To maximise the meter power detected, the polarizations of the two lasers were aligned orthogonally, allowing their separation after the fibre using polarizing beam splitters.
A diffraction grating further rejects any reflected signal beam from the input of the HC-PCF: this avoided saturation of the photodiode.

The $780$\,nm laser was designated as the \emph{meter} beam, and its phase shift was used to sense the power of the $776$\,nm \emph{signal} beam.
This choice resulted in  the strongest phase shift sensitivity; we note that signal and meter transitions are reversed when compared with that reported in Ref.\cite{Venkataraman2012aa}. 
An intermediate state detuning of $\Delta_i{\approx}1.2$\,GHz was used, along with low signal and meter powers to ensure that the atomic population in states $|i\rangle$ and $|e\rangle$ were minimised.
These measures ensured that the cross-Kerr effect was the dominant cause of the observed phase shifts.

%----------------------% Some Theory %----------------------%

The magnitude of the phase shift induced in the meter by the Kerr coupling can be characterised in three different ways: the meter beam's total phase shift; phase shift per photon; or the phase shift per atom.
For this excitation scheme the vapour's cross-Kerr coefficient takes the form $n_2\propto\rho\,\sigma_{met}\sigma_{sig} \lambda_{met}\lambda_{sig}$ \cite{Schmidt96}, where $\sigma$ and $\lambda$ are the atomic cross-sections and transition wavelengths respectively and $\rho$ is the atomic density.
It follows that the meter's total phase shift, $\phi_\mathrm{met}$, takes the form:
\begin{equation} 
	\begin{aligned}
	\phi_\mathrm{met} 	= & \frac{2}{3}L\,n_2\,k_\mathrm{met}\,P_\mathrm{sig}/A 	\\
	 				 \propto & \frac{2}{3}L\,\rho\,\sigma_{met}\sigma_{sig} \lambda_{sig}\,P_\mathrm{sig}/A 
	\end{aligned}\label{Eqn1}
\end{equation}
where $L$ is the length of the vapour-filled fibre, $k_\mathrm{met}$ is the meter's wavevector, $P_\mathrm{sig}$ is the signal power and $A$ is the mode area.
Equation~\ref{Eqn1} shows that the cross-Kerr coupling depends on atomic density, $\rho$, together with the ratio $\sigma_{met}\sigma_{sig}/A$.
The interaction time for a signal photon is set by the atomic decay rate, $\Gamma_{i}$, thus the phase shift per photon, $\phi_{ph}$, is:
\begin{equation} \label{Eqn2}
	\phi_{ph} = \phi_\mathrm{met}\,\hbar\,\omega_\mathrm{sig}\,\Gamma_{i}/P_\mathrm{sig} %\propto  L\,n_2\,k_\mathrm{met}\,\hbar\,\omega_\mathrm{sig}\,\Gamma_{i}/A 
\end{equation}
Finally the phase shift per atom, $\phi_{atom}$, is:
\begin{equation} \label{Eqn3}
	\phi_{atom} = \phi_\mathrm{met}/(\rho\,L\,A) %\propto n_2\,k_\mathrm{met}\,P_\mathrm{sig}/(\rho\,A^2)
\end{equation}
Importantly it can be seen that---in the absence of atomic saturation---$\phi_\mathrm{met}$ does not depend on the meter beam power.

\begin{figure}[t]
	\includegraphics[width=\columnwidth]{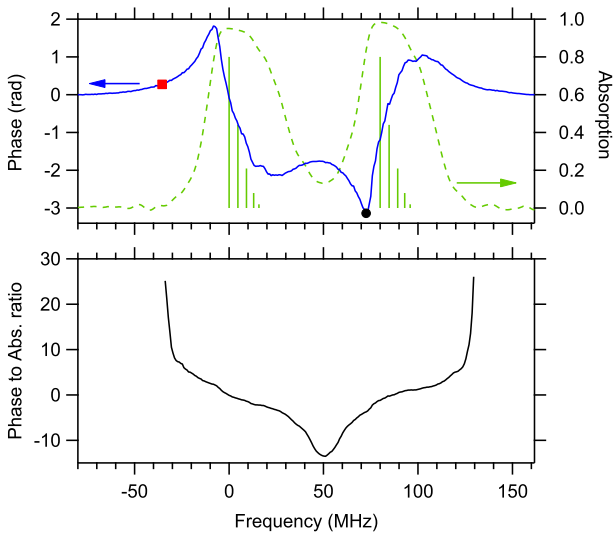}
	\caption{\label{Fig:PhaseSpectrum} Spectra of phase shift (blue) and absorption (green) as the two meter beams pass through the two-photon resonance (top).
	The ratio of phase to absorption is also shown (bottom, black).
	Optical powers were $P_{\textrm{met}} {\approx} 1$\,$\mu$W and $P_{\textrm{sig}} {\approx} 45$\,$\mu$W.
	Black circular and red square markers are referenced to in both Figs \ref{Fig:PhaseSummary} and \ref{Fig:MeterSat}.}
\end{figure}

To directly measure the phase shift induced by the signal beam, two separate meter beams of equal power, $P_{\textrm{met}}$, but different frequency, were coupled into the fibre.
The second meter beam was generated using an acoustic optic modulator (AOM) and was frequency offset by $80$\,MHz, see Fig~\ref{Fig:ExpSetup}c).
This frequency separation is larger than the transition manifold width, $\sim 32$\,MHz~\cite{Nez93}, which ensures that only one beam interacts with the transition at a time.
The non-interacting meter beam provided a phase reference while the second beam experiences the cross-Kerr phase shift.
A beat-note between the two meter beams was detected, both before, and after, the fibre, Fig~\ref{Fig:ExpSetup}\,c).
The former mixing product provided a RF phase reference which was compared to the output beat-note phase using an RF lockin amplifier.
This approach thus directly measures the cross-phase shift in the optical phase of the meter signal.
When compared to cross-phase measurements based on polarisation rotation~\cite{Venkataraman2012aa}, this approach is immune to unwanted birefringence changes in the fibre that may result from vibration or temperature changes generating both short and long term noise. Furthermore, this technique automatically rejects any self-phase modulation of the meter beam because the two beams composing the meter would suffer an equal phase shift.

%----------------------% Results %----------------------%

\begin{figure}[t]
	\includegraphics[width=\columnwidth]{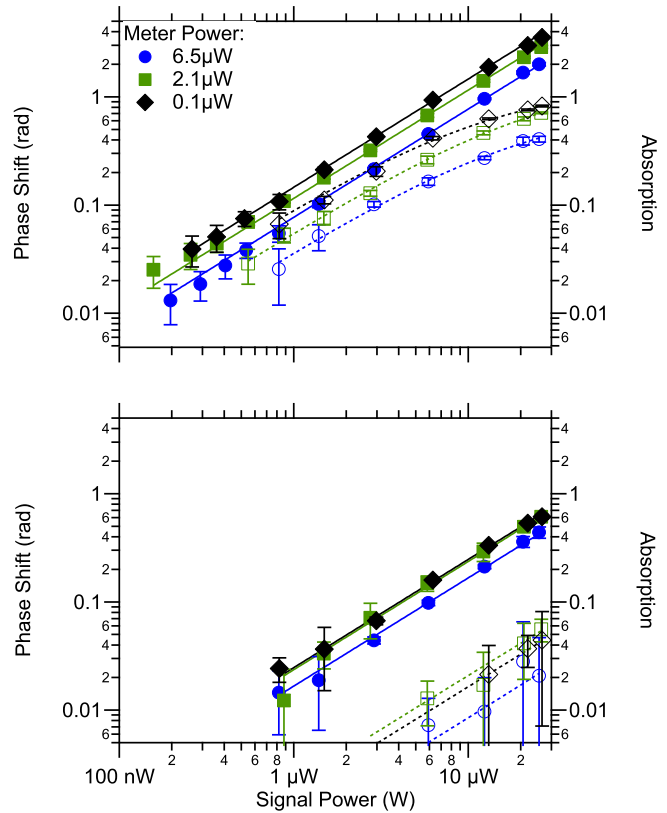}
	\caption{\label{Fig:PhaseSummary} Cross-phase shift (solid markers) and absorption (open markers) observed as a function of the signal power, $P_{\textrm{sig}}$, for an ensemble of meter powers.
Error bars represent the 95\% confidence interval.
Two situations are shown: \emph{top}, maximum recorded phase shift; and \emph{bottom} off-resonant phase shift, $\Delta_{e} {\approx} -35$\,MHz.
The panels are respectively at the detunings shown by the black-circular and red-square markers on Fig~\ref{Fig:PhaseSpectrum}.}
\end{figure}

A typical spectrum of the phase shift and absorption as the $780$\,nm laser was scanned through the two-photon transition is shown in the top panel of Fig~\ref{Fig:PhaseSpectrum}.
In this example a phase shift of up to $\pi$ radians was observed for $P_{\textrm{sig}}{\approx}25$\,$\mu$W and $P_{\textrm{met}}{\approx}1$\,$\mu$W.
Asymmetry in the measured phase shift arises from the asymmetric absorption profile due to the individual excited state hyperfine components, whose positions and absorption strengths~\cite{Nez93} are marked by vertical lines in Fig.~\ref{Fig:PhaseSpectrum}.
The bottom panel shows the ratio between phase shift and absorption which is found to increase with increasing $|\Delta_{e}|$, as expected from a two-level atomic model \cite{Rand2010}.
It is clear that operation at high detunings from the two-photon resonance can deliver reasonable phase shifts with exceedingly small absorption.

The sensitivity of the cross-phase modulation to both signal and meter powers was explored by varying each by over two orders of magnitude.
In each measurement, $5$ to $10$ spectra were taken in order to reduce statistical uncertainty on the measured phase shift.
For each spectra recorded, the measured dispersion curve was fitted and the phase shift calculated from this fit.

The top panel of Figure~\ref{Fig:PhaseSummary} shows the maximum phase shift---located at the point indicated by the black circle on Fig~\ref{Fig:PhaseSpectrum}---at various combinations of the signal and meter powers.
In contrast, the bottom panel shows the phase shift for an off-resonance signal where the absorption is strongly reduced, indicated by the red square on Fig~\ref{Fig:PhaseSpectrum}, $\Delta_{e}{\approx}-35$\,MHz.
At this point, the cross-phase shift is a factor of $\sim6$ times smaller than the maximum phase shift shown in the top panel, but the absorption is suppressed by more than a factor of ${\approx}20$.
Further detuning of $\Delta_{e}$ reduced the absorption below detectable levels for this experiment.

\begin{figure}[t]
	\includegraphics[width=\columnwidth]{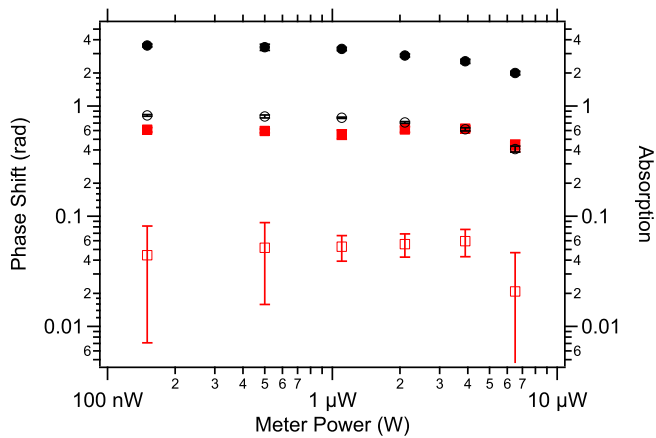}
	\caption{\label{Fig:MeterSat}  Phase shift (solid) and absorption (hollow) saturation as a function of meter power, $P_{\textrm{met}}$, for $P_{sig} {=} 25$\,$\mu$W.
	Saturation begins respectively at $P_{\textrm{met}}\approx 3$\,$\mu$W and $P_\mathrm{met} \approx 20$\,$\mu$W for the maximum-phase (black circle) and off-resonance (red square) cases.}
\end{figure}

We see from Fig~\ref{Fig:PhaseSummary} that across the full range of tested signal powers, our experimental results agree with Eqn.~\ref{Eqn1}, which predict $\phi_\mathrm{met} \propto P_\mathrm{sig}$, for a given meter power.
This agreement indicates that the Rb vapour is producing a classical Kerr phase shift.

Figure \ref{Fig:MeterSat} shows that the converse does not apply.
The phase shift measured for a given signal power is not independent of the meter power, as atomic saturation and population pumping effects begin at large $P_{met}$.
In the maximum-phase and detuned cases --- black circles and red squares respectively in Figure \ref{Fig:MeterSat} --- we see saturation begin at $P_{\textrm{met}}\approx 3$\,$\mu$W and $P_\mathrm{met} \approx 20$\,$\mu$W.
The saturation points are independent of signal power, as can be seen from the fact that the phase-shift lines remain parallel in the top panel of Figure \ref{Fig:PhaseSummary}, even when above the meter saturation power.

Knowing this, and that the the data from Fig.~\ref{Fig:PhaseSummary} shows an effective phase shift of $3.6$ rad for $P_{\textrm{sig}} {=} 25$\,$\mu$W, we use Eqn.~\ref{Eqn2} and \ref{Eqn3} and find phase shifts of $\phi_{ph} \approx 1.3 {\times} 10^{-6}$\,rad/photon and $\phi_{atom} \approx 2.9 {\times} 10^{-9}$\,rad/atom.
Such phase shifts correspond to a cross-Kerr, non-linear, index of $n_2=1.3{\times}10^{-6}$\,cm$^{2}$/W. These results are a factor of $10$ larger than that measured in non-linear glass waveguides~\cite{Matsuda2009, Matsuda2009APL}.

%----------------------% Shot Noise %----------------------%

The spectral density of the phase noise floor of our meter was $7 {\times} 10^{-5}/(\sqrt{P_\mathrm{met}/\mu\textrm{W}})$\,rad/$\sqrt{\textrm{Hz}}$ as directly measured at the output of the lockin amplifier measuring the meter.
This noise level was consistent with that calculated from the photon shot-noise of the meter beam, and its origin was verified by varying the meter power and observing the expected improvement in the sensitivity with the square-root of the power.
This sensitivity could be improved substantially by using a detector with a higher quantum efficiency for IR radiation than the one used here (4\%).

%----------------------% Conclusion %----------------------%

This work is a demonstration of the potential of this new platform for exhibiting strong photon-photon interaction while simultaneously showing low absorption.
%These attributes give the potential for this device  to be used in an optical cavity which can yield reliable shot-to-shot  measurements of the  signal photon number~\cite{fan2012breakdown}.
Furthermore, Eqn. \ref{Eqn1} and \ref{Eqn2} suggest several routes to improve performance.
Firstly reducing the core diameter to $5$\,$\mu$m improves atom-light coupling by a factor of ${\sim}80$. This has negligible effect on induced phase-shifts as long as the exciting optical pulses are shorter than the average transit-time for an atom across the fibre mode\cite{Venkataraman2012aa}. 
Second, the use of light-induced atomic desorption (LIAD) can increase the Rb density by a factor of $>200$ \cite{Slepkov2008, Bhagwat2009, Marmugi2012}, giving a consequent benefit in the cross-phase sensitivity.
A final factor can be gained through increasing the effective atom-light interaction length by a factor of 10. This can be achieved by either filling a longer length of HC-PCF, or using slow-light techniques \cite{Lukin2000}.
By using high quantum-efficiency detectors \cite{Smith2012aa} and the aforementioned techniques, the extrapolated sensitivity can approach $>0.2$\,rad/photon.
% at the single-photon level per atomic bandwidth ($10$\,pW), with a meter power of $10$\,$\mu$W.
In this regime we will be able to resolve the controversy between the predictions of the classical Kerr theory and the new quantum Kerr theory outlined in Ref.~\cite{fan2012breakdown}, and lay the foundation of a scalable photonic architecture for quantum information processing.

%Furthermore, a route for substantial enhancement of the performance can be seen by examining Eqns.~\ref{Eqn1} and \ref{Eqn2}.
%The use of smaller fibre cores \cite{Venkataraman2012aa} will enhance the atom-light coupling by a factor of around $\sim 100$ over that reported here. 
%A second enhancement factor can be gained by increasing the Rb density using techniques such as LIAD or by increasing the  temperature of the fibre \cite{Slepkov2008, Bhagwat2009, Marmugi2012}.
%We calculate that by using a high quantum-efficiency IR detector, together with the use of a fibre with a $6$\,$\mu$m core diameter, the experiment can operate at the single-photon detection level with an averaging time of just 10s.
%In this regime we can resolve the controversy between the predictions of the classical Kerr theory and the new quantum Kerr theory outlined in Ref.~\cite{fan2012breakdown}.

%----------------------% Acknowledgements %----------------------%
The authors acknowledge financial support from the Australian Research Council under grants DP0877938, DE120102028, FT0991631, and Centres of Excellence for Engineered Quantum Systems and Quantum Computing and Communication Technology.
AGW acknowledges support from the UQ Vice-Chancellor's Senior Research Fellowship.
We thank Eugene Ivanov for equipment loans.
We are grateful to Francois Couny for his contribution in the fibre fabrication the HC-PCF.
% and the Australian Microscopy and Microanalysis Research Facility at the University of Western Australia, a facility funded by the university and state and commonwealth governments.

\bibliography{XPS}

\end{document}